\pdfoutput=1

\documentclass[aps,prb,reprint,showpacs,superscriptaddress,byrevtex]{revtex4-1}

\usepackage{graphicx}
\usepackage{graphics}
\usepackage{amsmath}
\usepackage{amssymb}
\usepackage{amsfonts}
\usepackage{dcolumn}
\usepackage{dsfont}
\usepackage{latexsym}
\usepackage{rotating}
\usepackage{color}
\usepackage{latexsym}
\usepackage{bbm}
\usepackage{subfigure}
\usepackage{float}
\usepackage{epsfig}
\usepackage{epsf}
\usepackage{psfrag}
\usepackage{bm}
\usepackage{amsthm}
\usepackage{eucal}
\usepackage{mathrsfs}
\usepackage{url}
\usepackage{braket}
\usepackage{natbib}

\usepackage{color} 

\usepackage{hyperref}
\hypersetup{
colorlinks=true,final=true,
        linkcolor=blue,
        citecolor=blue,
        filecolor=blue,
        urlcolor=blue,
}

\begin{document}

\title{Anisotropic thermoelectric properties of EuCd$_{2}$As$_{2}$ : An Ab-initio study}
  
\author{Jyoti Krishna, Mukesh Sharma and T. Maitra}

\affiliation{ Department of Physics, Indian Institute of Technology Roorkee, Roorkee-247667, Uttarakhand,India}

 \begin{abstract}
In search of better thermoelectric materials, we have systematically investigated the thermoelectric properties of a 122 Zintl phase compound EuCd$_{2}$As$_{2}$ using \textit{ab-initio} density functional theory and semi-classical Boltzmann transport theory within constant relaxation time approximation. Considering the ground state magnetic structure which is A-type antiferromagnetic (A-AFM) and non-magnetic (NM) structure, we evaluated various thermoelectric parameters such as Seebeck coefficient, electrical and thermal conductivity, power factor and figure of merit (ZT) as function temperature as well as chemical potential. Almost all thermoelectric parameters show anisotropy between $xx$ and $zz$ directions which is stronger in case of A-AFM than in NM. Both A-AFM and NM phase of the compound display better thermoelectric performance when hole doped. We observed high Seebeck coefficient and low electronic thermal conductivity in A-AFM phase along $zz$ direction. The remarkably high ZT of 1.79 at 500 K in A-AFM phase and ZT$\sim$1 in NM phase suggest that EuCd$_{2}$As$_{2}$ is a viable thermoelectric material when p-doped.
\end{abstract}

		\maketitle
\section{Introduction}

The search for new and efficient theromoelectric(TE) materials has been one of the major goals of condensed matter, materials science and applied physics research in recent years for their energy related applications such as terrestrial cooling, recovery of heat waste etc\cite{bell,minnich}. Thermoelectric applications have also been utilized for long in space missions as thermoelectric power generators due to its reliability\cite{rowe}. Whether a particular material is suitable for TE device is governed by its figure of merit ($ZT$) (a measure of TE efficiency) which is a dimensionless parameter and is given by the following expression,
\begin{equation}
ZT = \frac{S^{2}\sigma T}{\kappa}
\end{equation}
Thus a good $ZT$ demands high Seebeck coefficient ($S$) and electrical conductivity ($\sigma$) and low thermal conductivity ($ \kappa = \kappa_{e}+\kappa_{l} $ where $\kappa_{e}$ and $\kappa_{l}$ are electronic and lattice contribution respectively). Further, the Seebeck coefficient ($S$) or thermopower given by Mott formula indicates  
that it is a measure of electronic structure asymmetry and rate of scattering near Fermi energy ($E_{F}$) which for a metallic or degenerate semiconductor reduces to:
\begin{equation}
S = (8\pi^{2}k_{B}^{2}/3eh^{2})m^{*}T(\pi/3n_{c})^{2/3}
\end{equation}
where, $k_{B}$ is the Boltzmann constant and e is the electronic charge. Thus $S$ solely depends on carrier density $n_{c}$ and effective mass $m^{*}$ of the charge carrier. As $S$ and $\sigma$ have opposite behavior i.e., the increase in Seebeck coefficient decreases the electrical conductivity and vice versa, the optimization of these two parameters for a good figure of merit becomes one of the challenging tasks. The only parameter we can control independently is $\kappa_{l}$ by choosing different crystal structure. Thus, a worthy choice of material is a pre-requisite\cite{susan}. 
\begin{figure}[ht!]
	\centering
	\includegraphics[width=8.5cm]{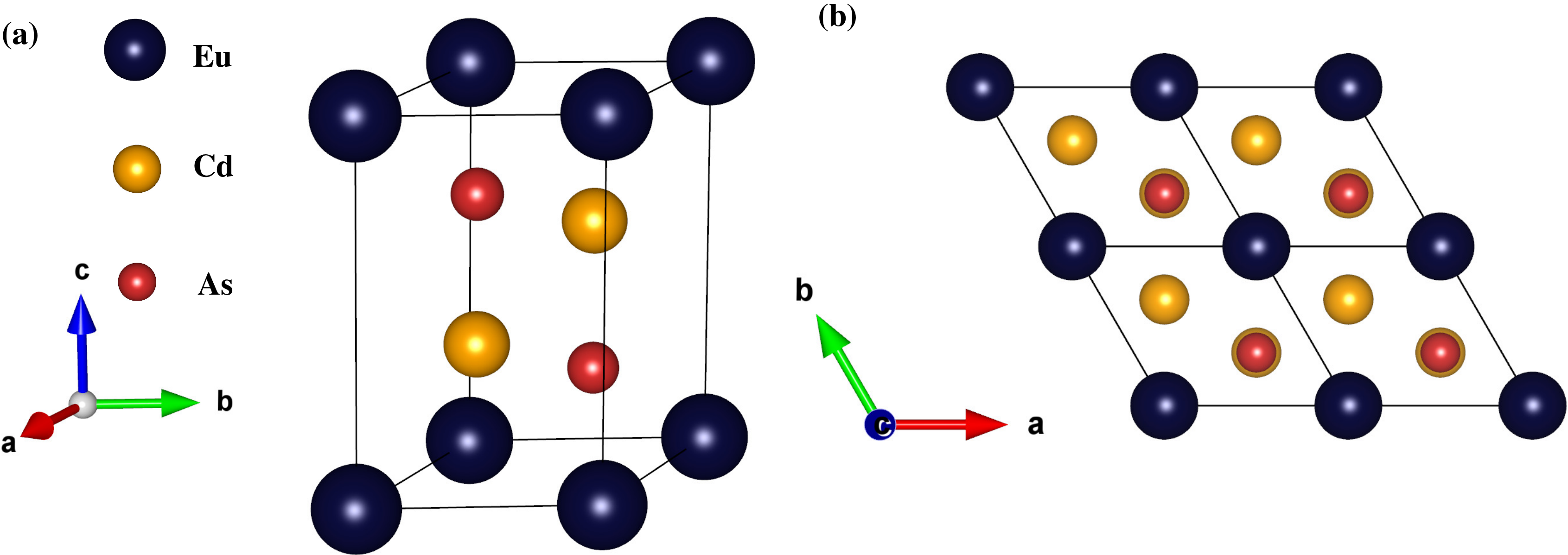}
	\caption{(a) The primitive unit cell of EuCd$_{2}$As$_{2}$ compound. (b) The crystal structure viewed from the top (c-axis) showing the hexagonal symmetry of the crystal structure.}
\end{figure}
The maximal $ZT$ has been found near the crossover region between semiconductors and metals\cite{Zheng}. The Zintl phase compounds come within this regime. Previous studies reported in literature proved these materials to be promising candidates for TE applications\cite{kauz_4,gascoin_4,west_4}. The crystal structure is also suitable for high $ZT$\cite{es,zeval,brown} values. In Zintl phases, there may be complete charge transfer from cations (group I or II, rare earth metals) to anionic slabs making it valence precise\cite{dokumen}. \\

Layered 122 type Zintl compound namely,  EuCd$_{2}$As$_{2}$  is one such compound which has trigonal crystal symmetry (see Fig.1) as in  CaAl$_{2}$Si$_{2} $ structure. Previous isostructural compounds have shown remarkable applications in TE\cite{min,fang}. For instance, at 700 K it has been reported that  EuZn$_{2}$Sb$_{2}$ and YbCd$_{1.6}$Zn$_{0.4}$Sb$_{2}$ has $ZT$ of 0.9\cite{hzhang} and greater than 1 \cite{wang} respectively. In  EuCd$_{2}$As$_{2}$, the layered structure of the compound can have an important role in lowering down the $\kappa$. Also, the strong covalent nature of polyanionic network of $[CdAs]^{2-}$ (due to similar electronegativity) can facilitate electron transport which is expected to increase the contribution of $\sigma$ in ZT. Most importantly, the strong scattering nature of Eu moments can effieciently scatter off the conduction electrons coming from $Cd$ and $As$ ions, thus minimizing the lattice effect. The high effective mass of Eu $ 4f $ electrons near Fermi level(FL) will also aid in enhancing $S$. It has been shown theoretically that if the density of states (DOS) approaches Dirac delta function (as for $ f $-electrons here), one can obtain a $ZT$ value as high as 14\cite{Zheng,mahan}. Considering highly peaked Eu DOS in  EuCd$_{2}$As$_{2}$ near FL\cite{jk_4} and the factors mentioned above, it is worth investigating  EuCd$_{2}$As$_{2}$ for the TE applications. From an earlier theoretical study\cite{jk_4} and experimental measurements (REXS)\cite{rahn}, it is established that the ground state magnetic order in  EuCd$_{2}$As$_{2}$ is A-type Antiferromagnetic (A-AFM) and the high temperature phase is paramagnetic. Therefore, in this work we investigate TE properties of both magnetic (A-AFM) and non-magnetic (NM) phase using density functional theory (DFT) calculations.  
\section{Methodology}
The calculations were performed using two density functional theory (DFT) codes: (1) the plane-wave pseudopotential based method as implemented in Vienna Ab-Initio Simulation Package (VASP)\cite{vasp_4} and (2) the full-potential linear augmented plane-wave (FP-LAPW) method as implemented in WIEN2k\cite{wien_4}. The structural details have been taken from experiment\cite{schell}. First we performed geometrical optimization of the structure with A-AFM and NM state using the VASP code. For this, we employed PBE-GGA\cite{perdue_4} as exchange-correlation functional and PAW method for pseudopotential\cite{andrew}. We used plane wave basis set with kinetic energy cut off of 400 eV and the k-point mesh of $8\times8\times8$ over the full Brillioun zone was used to achieve the force convergence upto $10^{-8}$ $eV/\AA$ as the atomic positions must be optimized properly to ensure that there is no residual force at T= 0 K. This optimized structure was then further checked for its dynamical stability using PHONOPY\cite{phonopy_4} package. For this, we have considered a supercell of size $2\times2\times1$. Once all the stability criteria have been fulfilled by both the magnetic structures, the final structures are used for the calculation of the TE properties. The further calculations were performed using the WIEN2k code within Perdew-Burke-Ernzerhof generalized gradient approximation (PBE-GGA)\cite{perdue_4} as an exchange-correlation functional. Since the system is semimetallic, we took 896 $\vec{k}$ points in the irreducible Brillouin zone for the proper convergence of energy eigenvalues. The plane wave cutoff parameter (R$_{mt}$K$_{max}$) was set to 7.0 for all the calculations. The radii separating core and valence electrons, i.e., muffin-tin radius is 2.5 a.u. for $Eu$/$Cd$ and 2.49 a.u. for $As$ and we considered the plane wave expansion in spherical harmonics up to angular momentum quantum number $l$ = 10. The energy eigenvalues obtained in denser k-mesh are now used to evaluate the thermoelectric properties. This was carried out within semi-classical Boltzmann transport theory as implemented in BoltzTraP\cite{boltz_4} code.

\begin{figure*}[!ht]
	\centering
	\includegraphics[scale=0.5]{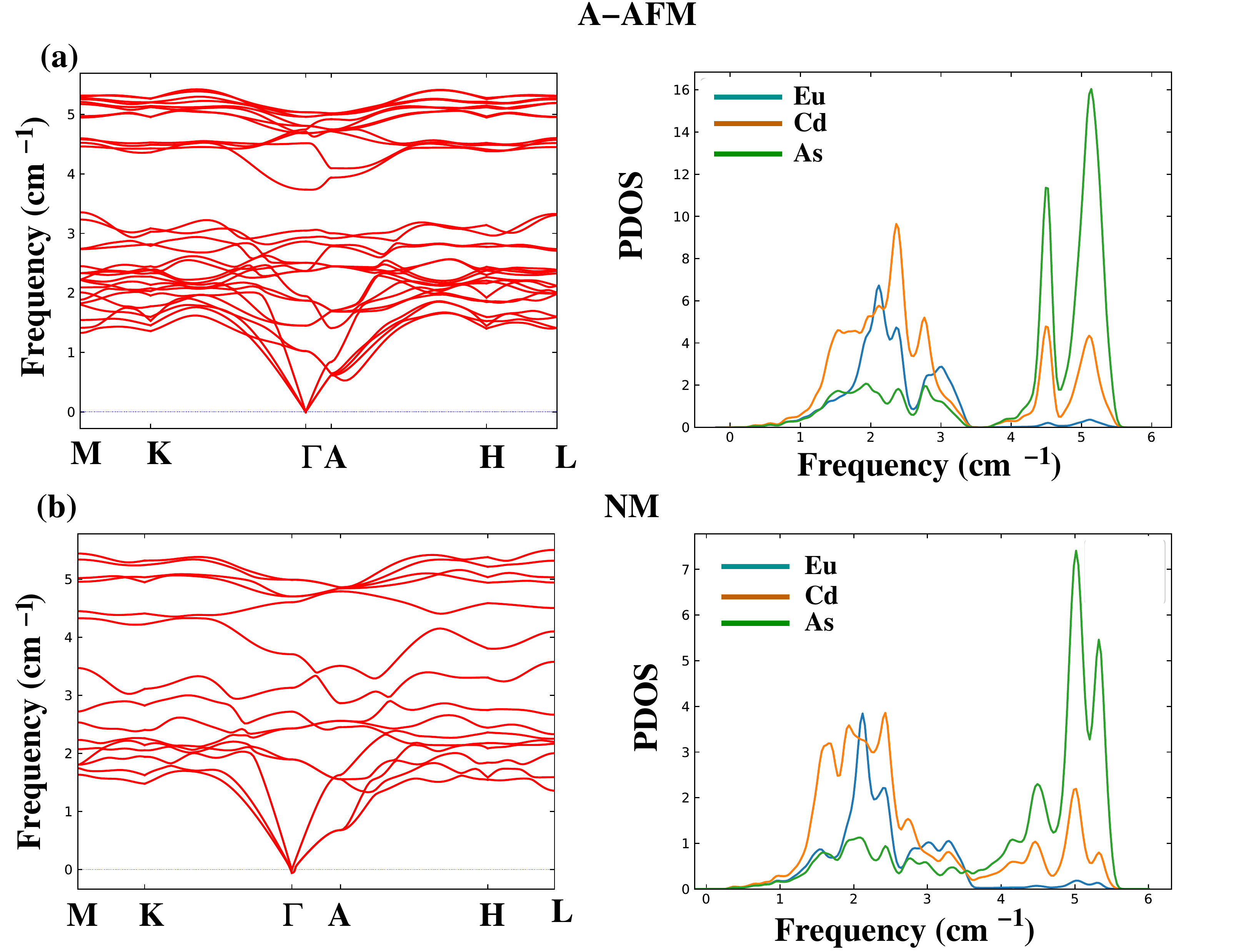}
	\caption{Phonon dispersion along high symmetry path in the hexagonal Brillouin zone and phonon partial density of states (PDOS) for A-type AFM and NM magnetic structures.}
\end{figure*}

\section{Results and Discussions}
\subsection{Structural Optimization and Dynamical Stability}

Before proceeding directly to obtain the TE parameters, first the geometrical and dynamical stability of the structure must be ensured. For this, we have first optimized the crystal structure obtained experimentally\cite{schell} for A-AFM and NM configurations by minimizing the Hellmann-Feynmann force between atoms. Since, the Eu ions are heavier only atomic positions of Cd and As deviated by about 0.1 \% and 2.4 \% respectively from the experimental atomic positions. The optimized and the experimental bond lengths are shown in Table I. The calculated structure is in good agreement with the experiment.

\begin{table}[!ht]
	\centering
	\caption{Experimental and calculated bond lengths (in $\AA$) for  EuCd$_{2}$As$_{2}$.}
	\
	\renewcommand{\arraystretch}{1.5}
	\label{tab1}
	\scalebox{0.9}{
		\begin{tabular}{||c|c|c||}
			\hline
			& \textbf{Exp.} & \textbf{Calculations}  \\
			\hline\hline
			\textbf{Cd-As (along ab plane)} & 2.717  & 2.7382  \\
			\hline
			\textbf{Cd-As (along c-axis)} & 2.841 & 2.941  \\
			\hline
			\textbf{Cd-Eu} & 3.724 & 3.7112  \\
			\hline
			\textbf{As-Eu} & 3.145 & 3.097  \\
			\hline\hline
	\end{tabular}}
\end{table}

\begin{figure}[ht!]
	\centering
	\includegraphics[scale=0.25]{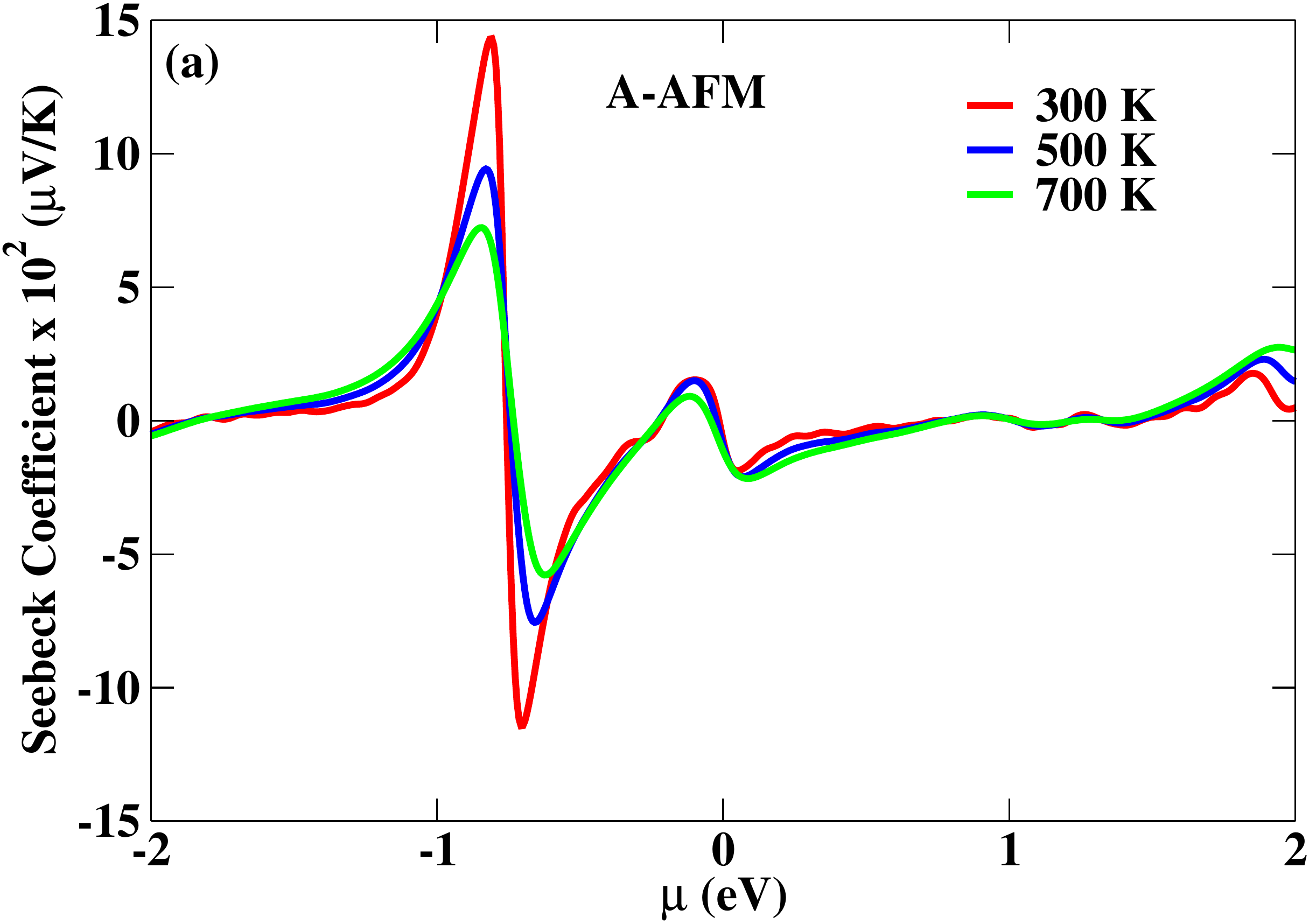}
	\includegraphics[scale=0.25]{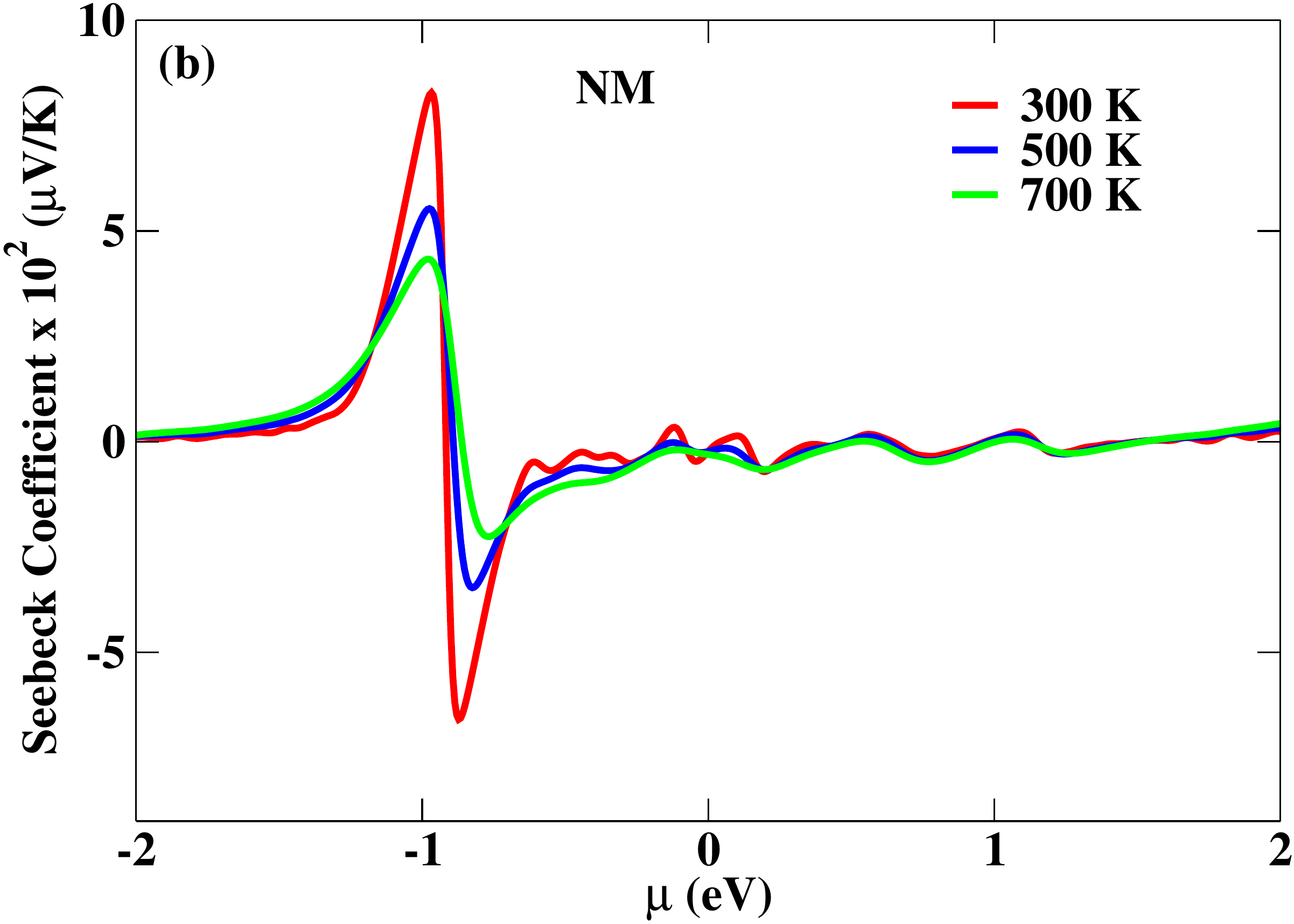}
	\includegraphics[scale=0.25]{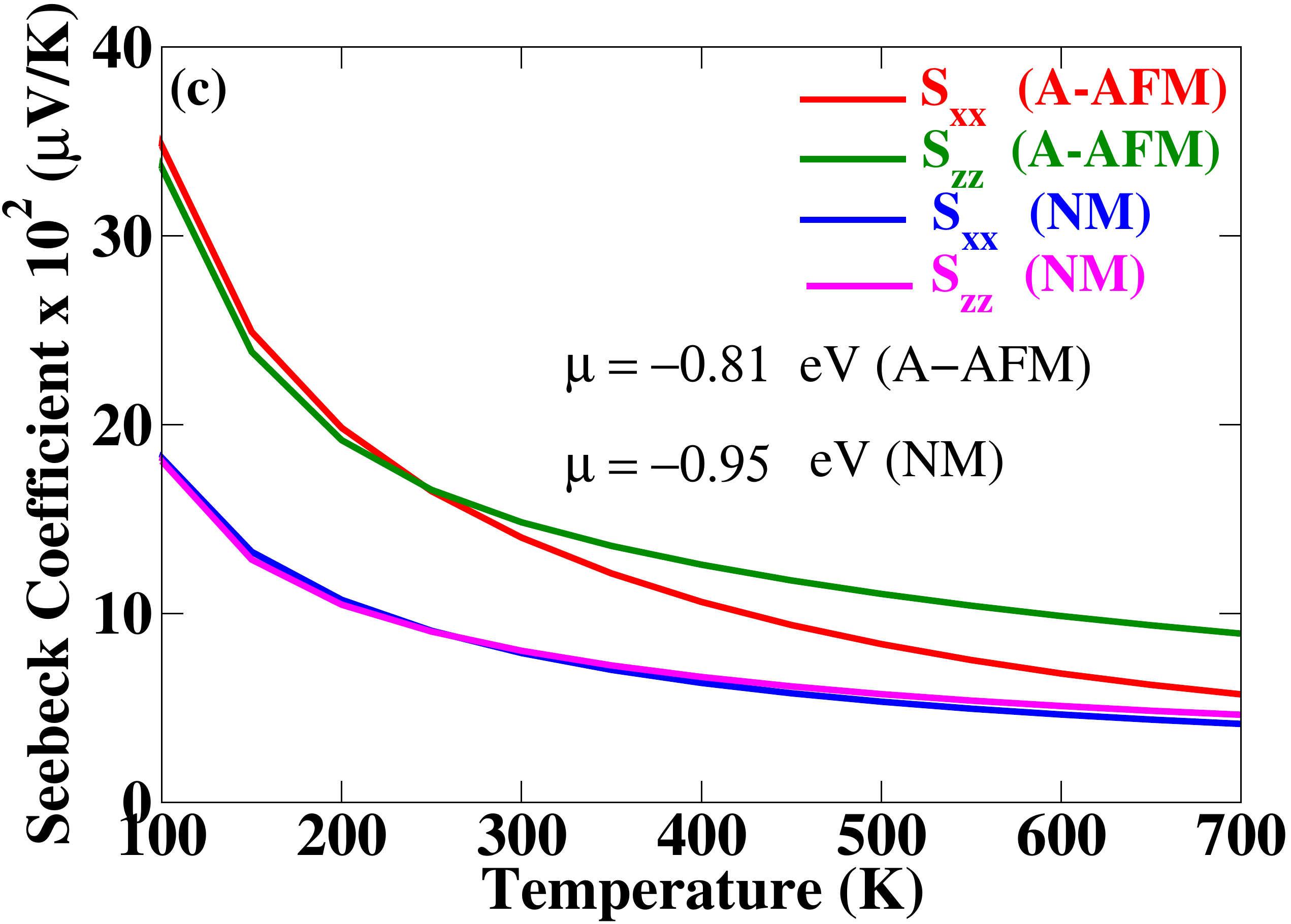}
	\caption{Variation of Seebeck coefficient (S) as a function of chemical potential ($\mu$) for (a) A-type AFM, and (b) non-magnetic (NM) state showing prominent contribution of $S$ when p-doped, (c) $S$ as a function of temperature along $xx$ and $zz$ direction for both structures for $\mu$=-0.81 eV and -0.95 eV for A-AFM and NM respectively. The Fermi level is set at $\mu$= 0 eV.}
	\label{fig1_4}
\end{figure}
Once the optimized structure is obtained, the crystal structures are further checked for their dynamical stability using phonon calculations within finite displacement approach\cite{parlinski_4,phonopy_4}. Within harmonic approximation, only real and positive phonon frequencies represent the stability of the system. Fig. 2 represents the phonon bandstructure and partial density of states (PDOS) for NM and A-AFM cases. In NM, the number of atoms present in primitive unit cell is N=5. Thus, 3N=15 phonon branches are there. Out of these branches, bottom three branches represent the acoustic modes and remaining 12 represent the optical modes. On the other hand, in A-AFM, the unit cell gets doubled hence total 30 phonon branches are there out of which 27 branches are the optical branches. In both the cases, acoustic branches are divided into two transverse and one longitudinal mode, which have linear dispersion close to the $\Gamma$ point. We can see that in NM case, for frequencies below 2 $cm^{-1}$, the lower lying optical branches intersect with the acoustic branches and in A-AFM case, the intersection is below 0.6 $cm^{-1}$ which indicates strong phonon-phonon scattering in the system. This strong scattering decreases the mean free path hence limits the thermal conductivity which is indeed a good signature for the better TE performance\cite{anoop}. In NM case we observed very small negative frequency (imaginary frequency) of -0.07 $cm^{-1}$ at $\Gamma$ point which is not of much concern, because this can be eliminated if we use anharmonic approximations. Whereas, for A-AFM, no imaginary frequencies have been observed. Hence we confirm that both the magnetic phases of  EuCd$_{2}$As$_{2}$ are in equilibrium. From the phonon PDOS we observed strongly coupled vibrations between Eu,  Cd  and  As  ions. Since the  As  ion is lighter, it has highest peak at frequency 5.13 $cm^{-1}$ and 5 $cm^{-1}$ for A-AFM and NM case respectively. Besides, we can also observe a phonon gap at $\sim 3.5 cm^{-1}$ in A-AFM which is again because of low lattice thermal conductivity\cite{shuang} in this whereas no such gap has been observed in NM case.	

\begin{figure}[ht!]
	\centering
	\includegraphics[scale=0.3]{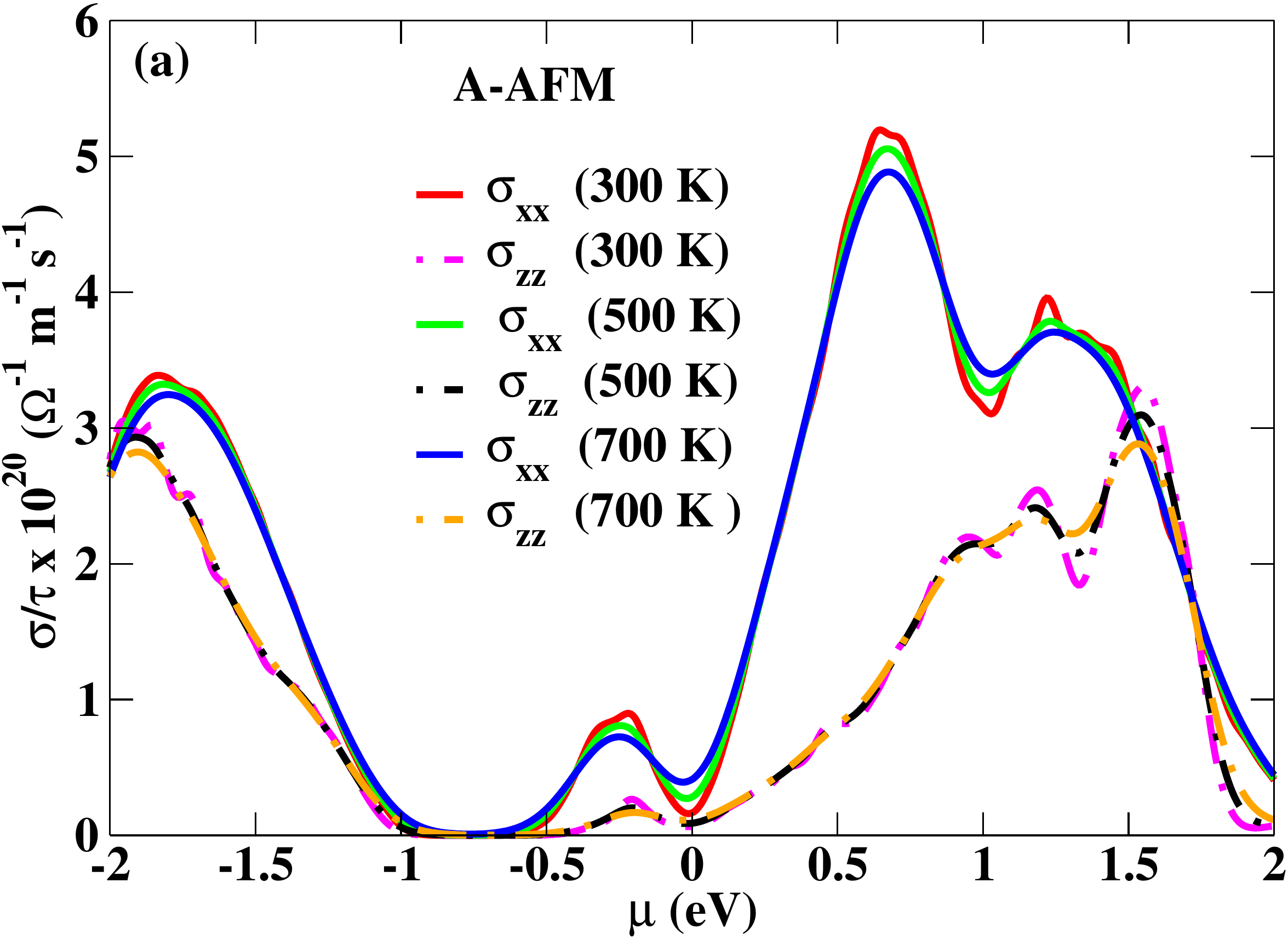}
	      \includegraphics[scale=0.3]{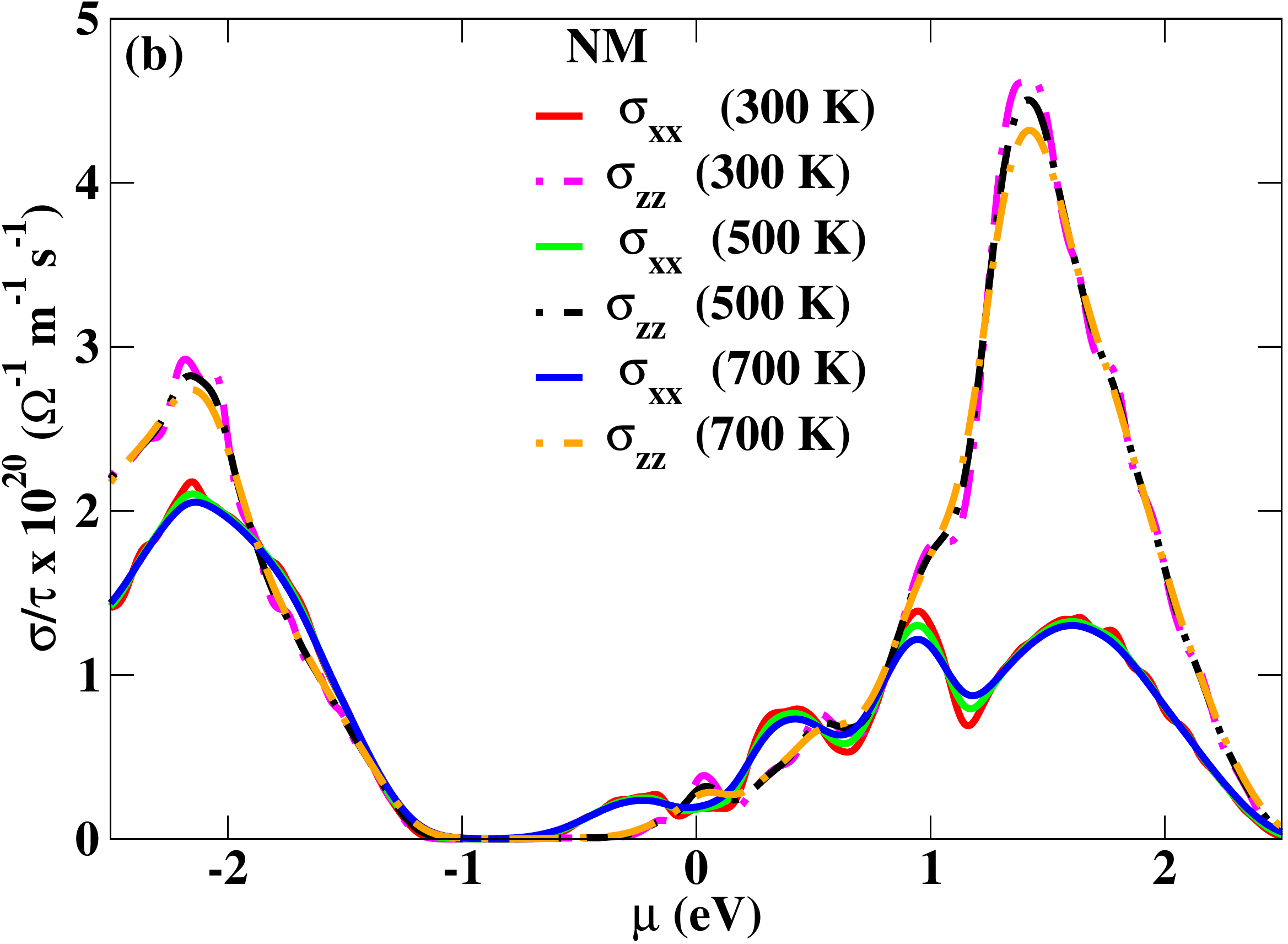}
	\caption{Electrical conductivity ($\sigma/\tau$) profile w.r.t chemical potential ($\mu$) for A-AFM and NM state along $xx$ and $zz$ direction for various temperatures. The Fermi level is set at $\mu$= 0.}
\end{figure}

\subsection{Thermoelectric Properties}
For the transport properties' calculations rigid band approximation (RBA) and constant relaxation time approximation (RTA) were employed. Under RBA it is assumed that with doping the FL can shift up or down without influencing the bandstructure. But this approximation is valid for low doping regime only. On the other hand, constant RTA assumes that the electron's relaxation time $\tau_{n\vec{k}}$ is independent of both band index (n) and k-point ($\vec{k}$) ($\tau_{n\vec{k}} = \tau $) and can be chosen by fitting with experimental results. Using the optimized structures obtained, we studied the thermoelectric properties as a function of chemical potential ($\mu$) and temperature (T) for both A-AFM and NM. In Fig. 3, we present the Seebeck coefficient ($S$) as a function of chemical potential ($\mu$). From $S$ versus $\mu$ graph we observe that peaks in $S$ appear just below FL ($\mu$= 0 eV) both on positive and negative side which indicates that with hole doping, the system can have better thermoelectric performance. Also, the larger magnitude of positive peak than the negative one indicates the dominant contribution of p-type charge carriers. At 300 K, the heights of two peaks in AFM case are 1423 $\mu$V/K and -1150 $\mu$V/K at $\mu$ -0.81 eV and -0.7 eV respectively. Whereas in NM the peak heights are 825 $\mu$V/K and -665 $\mu$V/K at -0.95 eV and -0.85 eV respectively. Also, it is observed that the overall Seebeck coefficient for the NM case is relatively smaller than that in A-AFM. This is because of the presence large number of states near FL in NM case while A-AFM is a low density semimetal\cite{jk_4,hpwang}. Further, in case of NM, $S$ peak has shifted much below FL with respect to A-AFM with the peak at $\mu$ value -0.81 eV (A-AFM) and -0.96 eV (NM) at 500 K. As the temperature increases, the magnitude of $S$ diminishes in both cases. The reason for this is broadening of Fermi distribution due to thermal excitation of the minority charge carriers that enhances the charge transport which ultimately reduces $S$.  
\begin{figure*}
	\centering
	\includegraphics[scale=0.3]{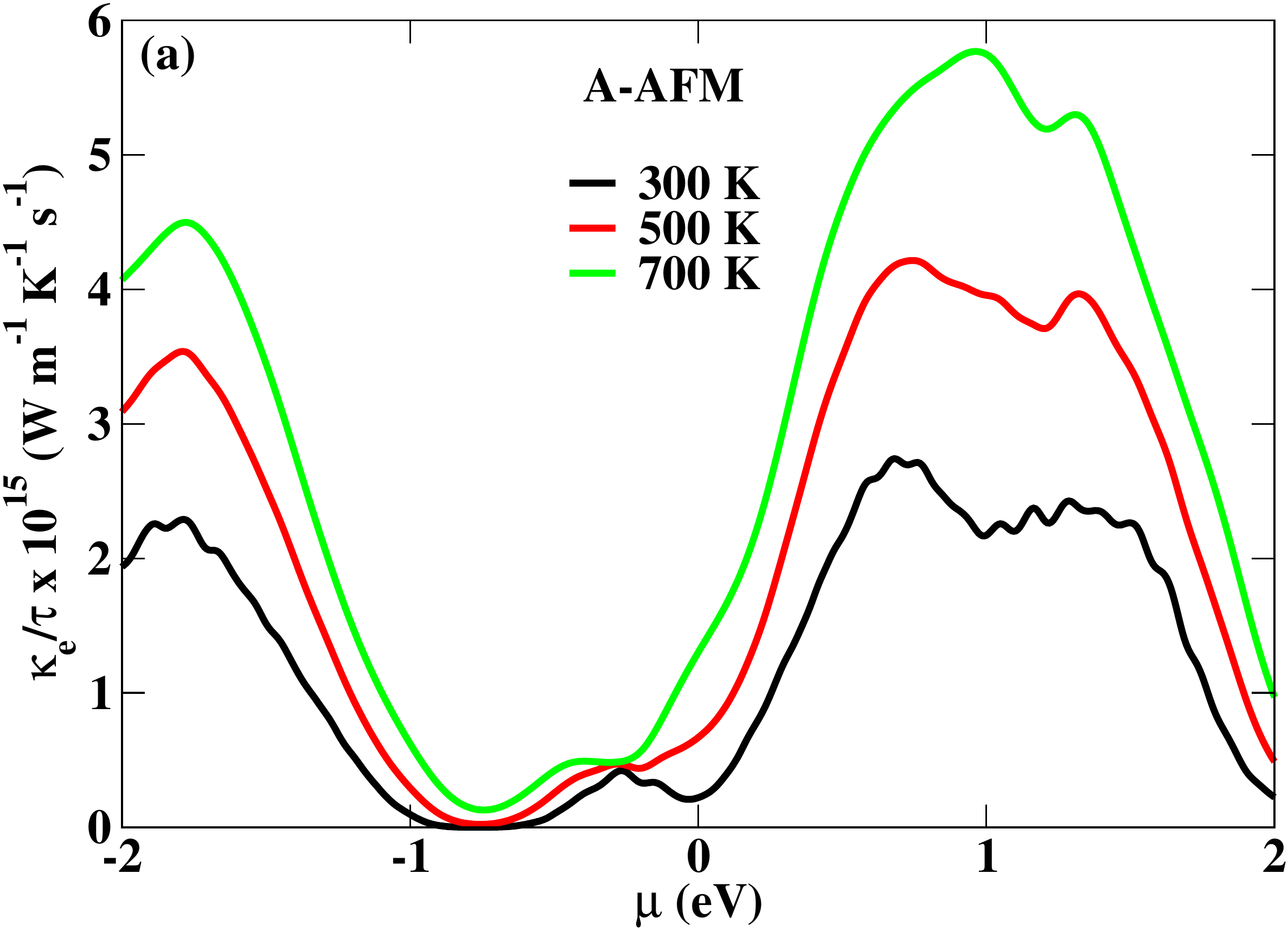}\includegraphics[scale=0.3]{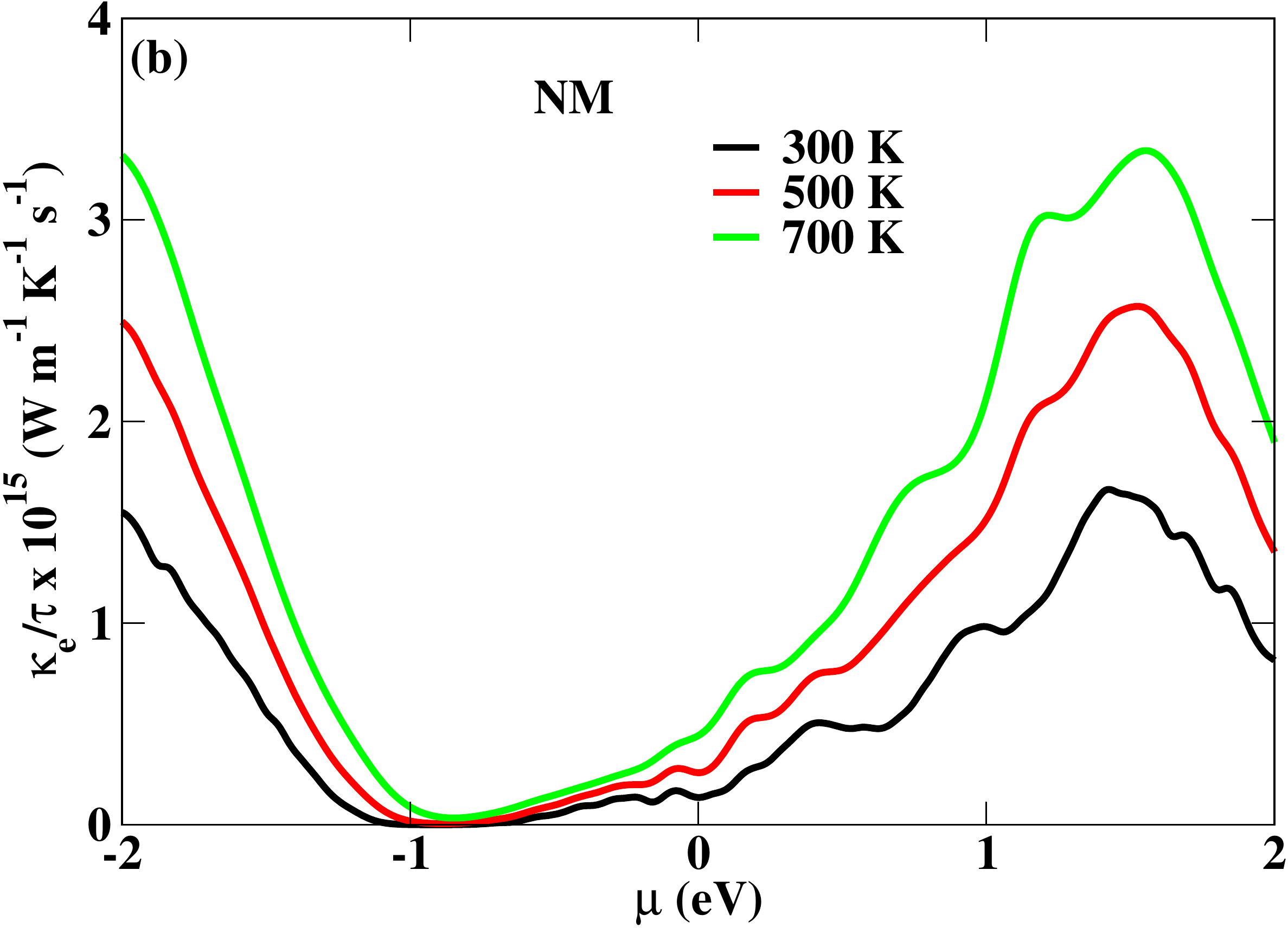}
	\includegraphics[scale=0.3]{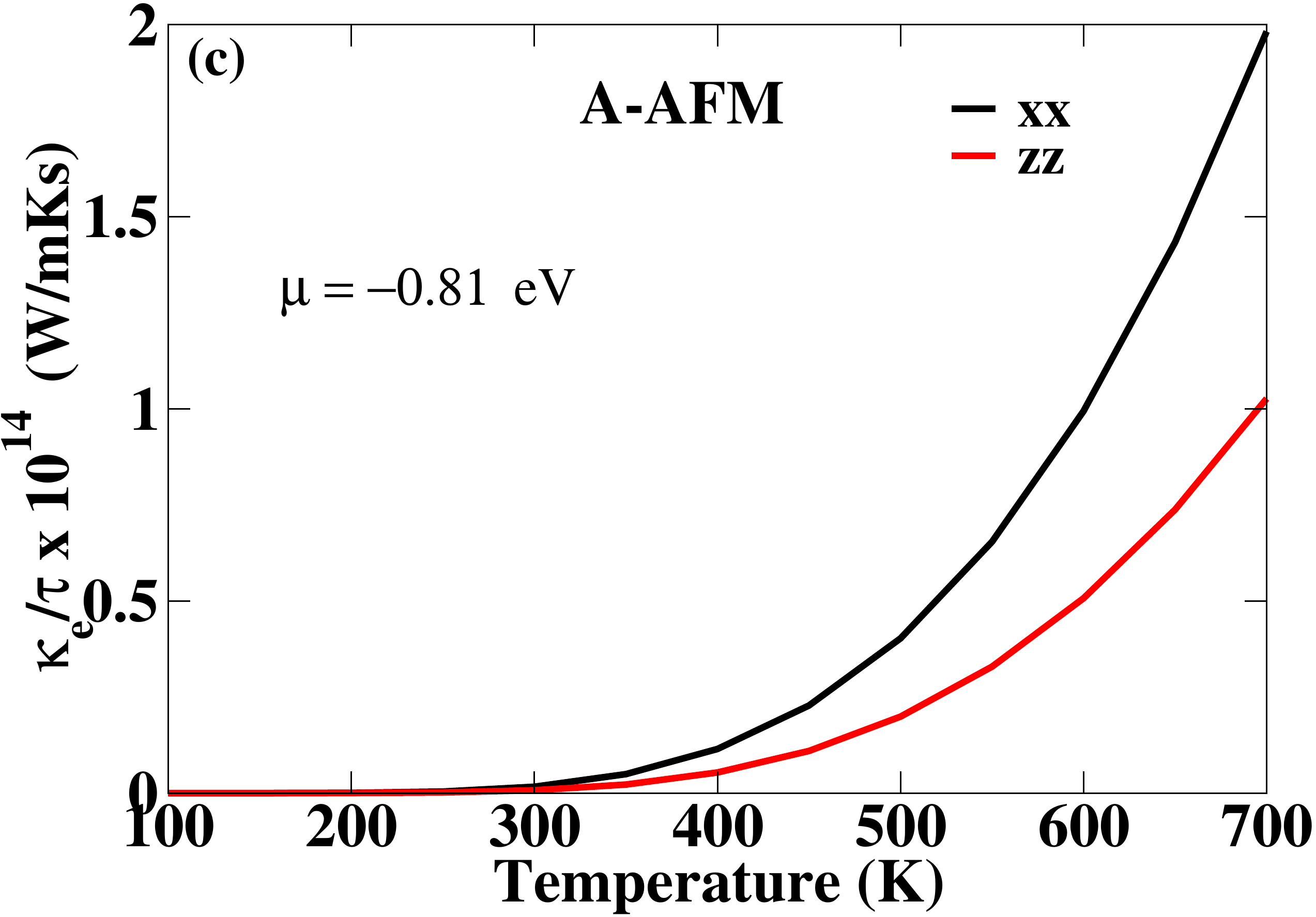}\includegraphics[scale=0.3]{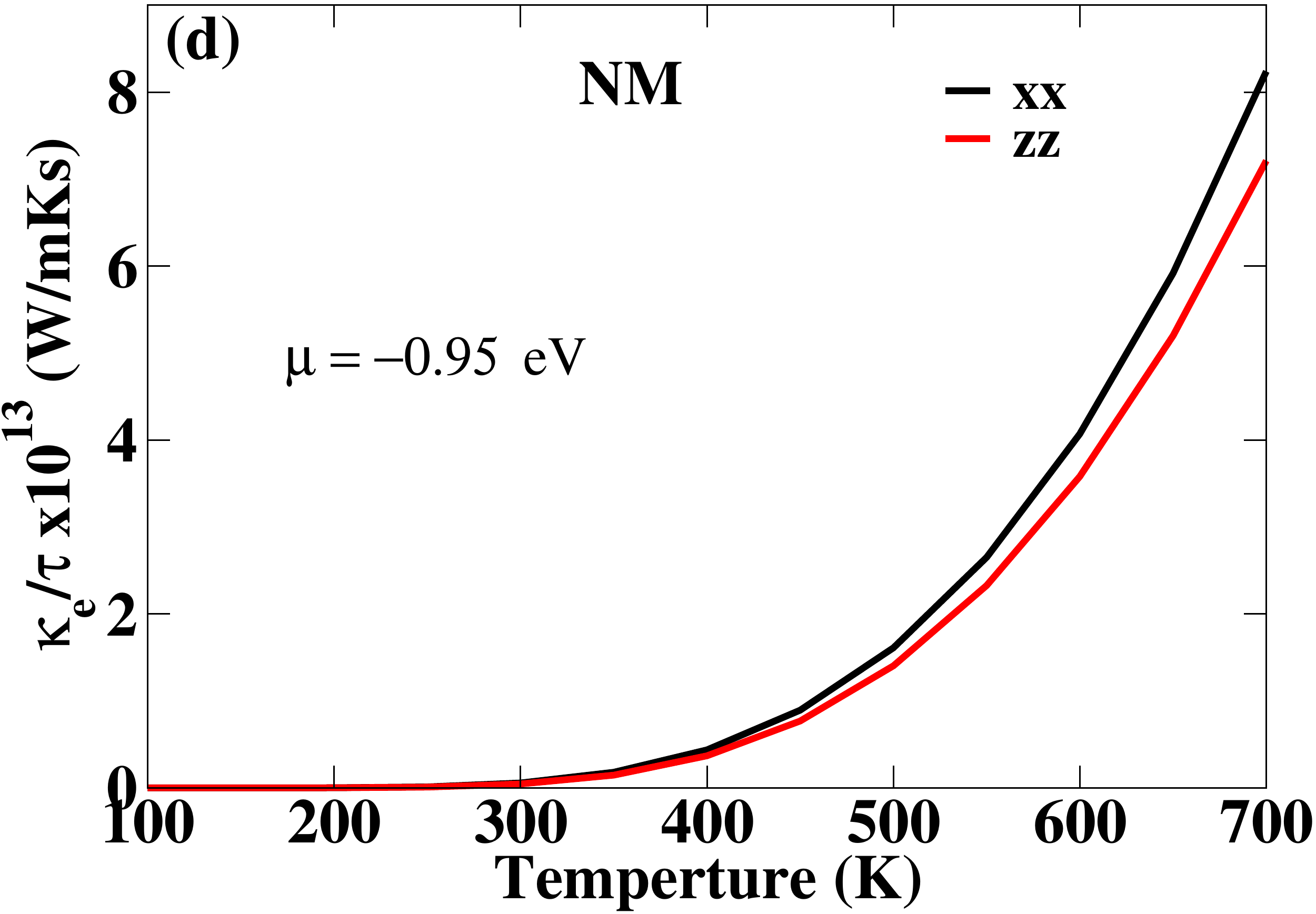}
	\caption{Variation of electronic thermal conductivity $\kappa_{e}/\tau$ w.r.t. $\mu$ for (a) A-AFM, and (b) NM state. The Fermi level is set at $\mu$= 0. (c) and (d) represent temperature dependence of $\kappa_{e}/\tau$ along $xx$ and $zz$ direction for $\mu$ = -0.81 eV and -0.95 eV for A-AFM and NM phases respectively.}
\end{figure*}
In an earlier study on the compound\cite{jk_4}, it has been observed that the  EuCd$_{2}$As$_{2}$ compound shows anisotropy in transport and magnetic properties. This anisotropy can be seen in the thermoelectric properties as well. Fig. 3(c) shows different contributions of Seebeck coefficient for $xx$ and $zz$  directions in both the A-AFM and NM cases. When the $\mu$ is fixed to the value at which we obtained high $S$ (0.81 eV and 0.95 eV below FL for A-AFM and NM respectively), we observed decrease in magnitude of $S$ with an increase in temperature as observed before. The value of $S_{xx}$ is lesser as compared to $S_{zz}$ at higher temperatures because of low carrier mobility along $z$ direction. This difference in $xx$ and $zz$ directions becomes significant in A-AFM case when temperature crosses 250 K whereas no large difference is seen in NM case. This could be again because of the minority charge carriers transport along all directions in NM. At 500 K the magnitude of $S_{zz}$ for A-AFM and NM is 1104 $\mu$V/K and 573 $\mu$V/K respectively because of the reason stated above. Thus we observe highest Seebeck Coefficient in EuCd$_2$As$_2$ in A-AFM along $zz$ direction. \\
The variation in electrical conductivity ($\sigma$/$\tau$) as a function of temperature has already been investigated \cite{jk_4} where metallic phase is observed at higher temperatures with large anisotropy (i.e.lower electrical conductivity along $zz$ direction as compared to $xx$) at lower temperatures. Here, we calculate the $\sigma$ as a function of $\mu$ along $xx$ and $zz$ direction for A-AFM and NM which is presented in Fig. 4. We see a clear anisotropy between $\sigma_{xx}$ and $\sigma_{zz}$. In A-AFM case the magnitude of $\sigma_{xx}$ and $\sigma_{zz}$ are higher in the region $\mu>0$ which implies higher conductivity when the system is doped with electrons. The overall contribution of $\sigma$ along $xx$ (peak value of 5.186$\times$$10^{20}$ at $\mu$ = 0.64 eV) is found to be larger as compared to that along $zz$ (peak value of 3.274$\times$$10^{20}$ at $\mu$ = 1.55 eV). This is related to the higher resistivity along $zz$ direction. However, in the NM case $\sigma_{zz}$ has pronounced peaks around $\mu =1.5 eV$ and $\mu =-2eV$. \\
\begin{figure*}[ht!]
	\centering
	\includegraphics[scale=0.3]{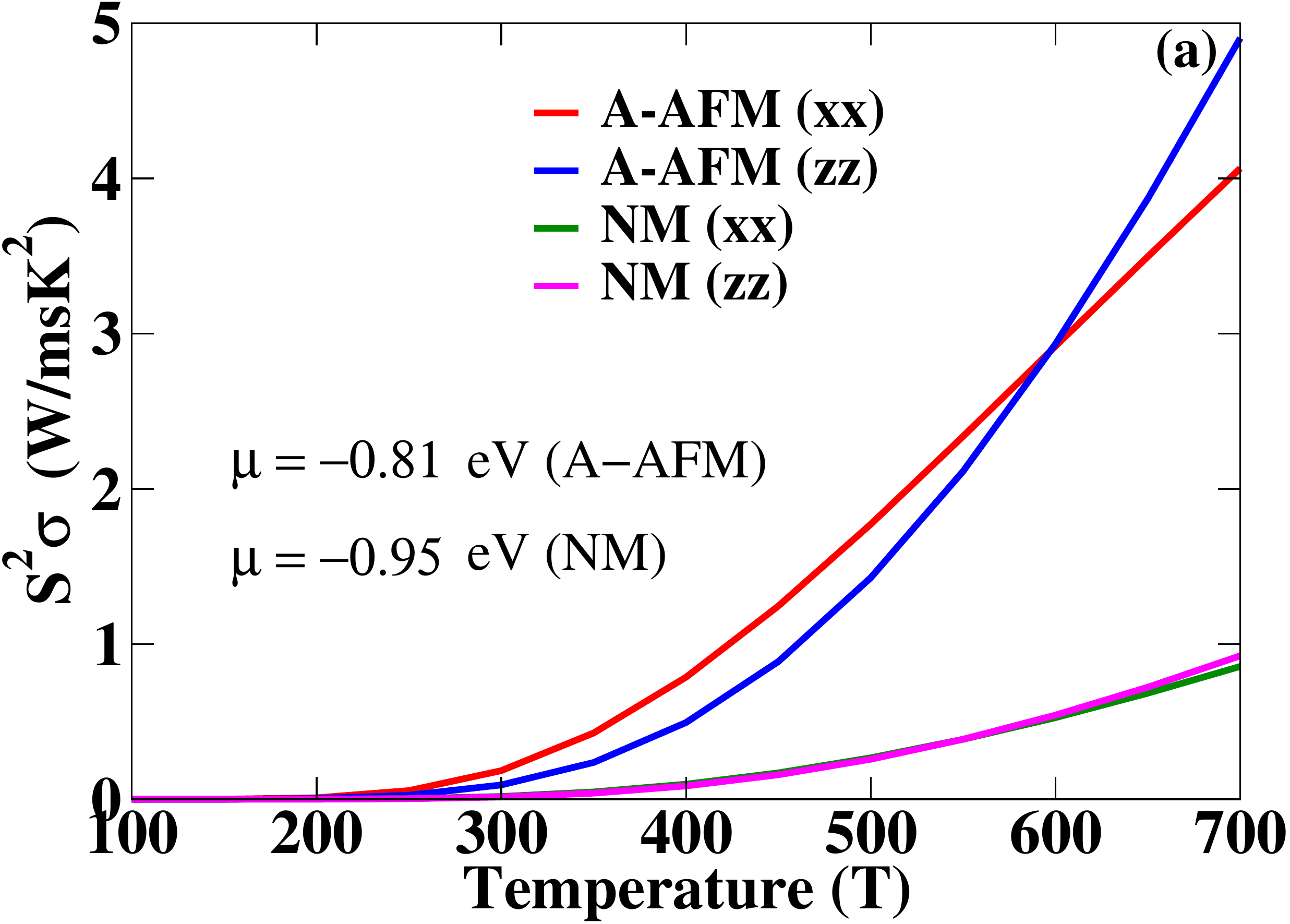}\includegraphics[scale=0.3]{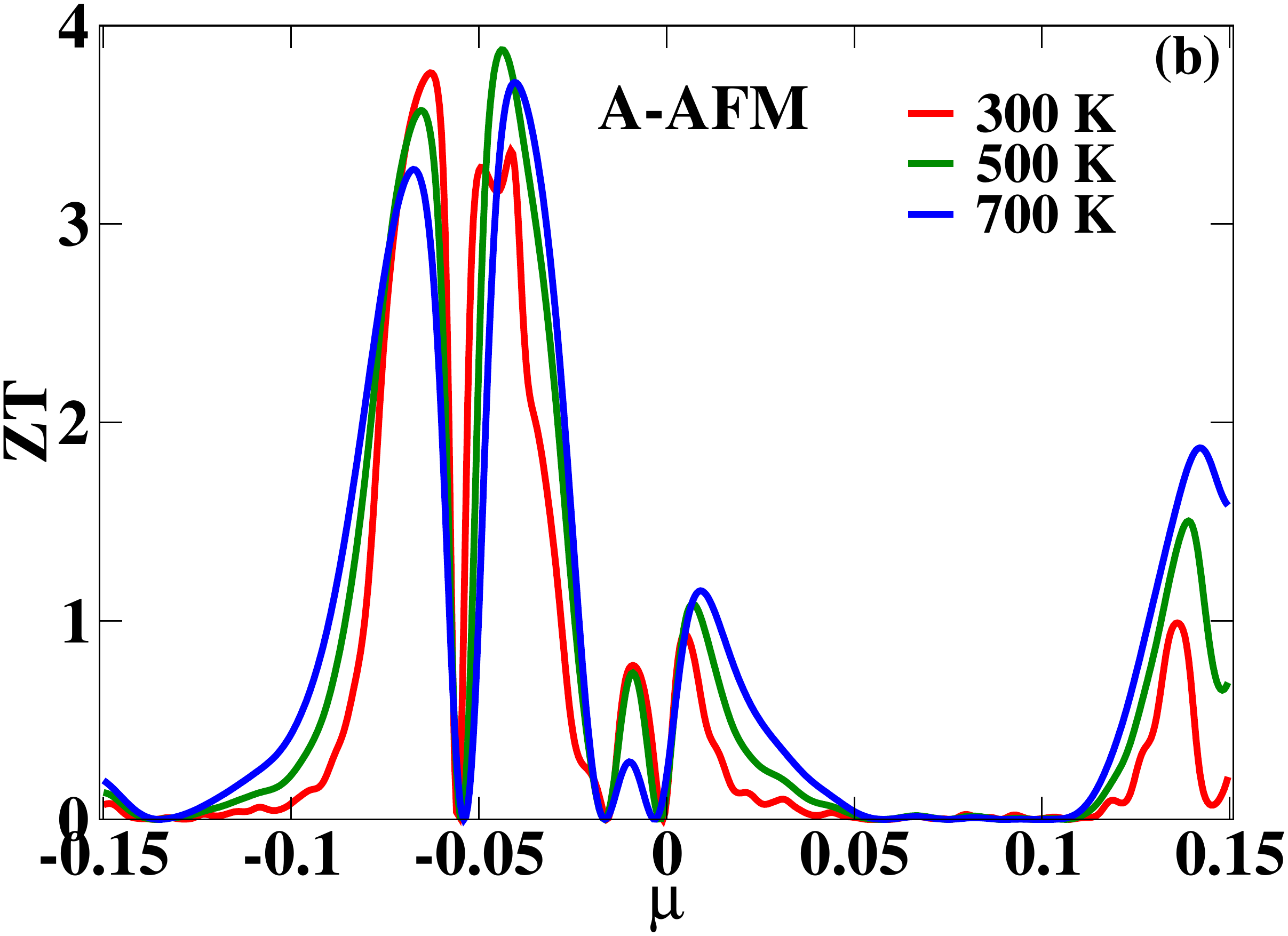}
        \includegraphics[scale=0.3]{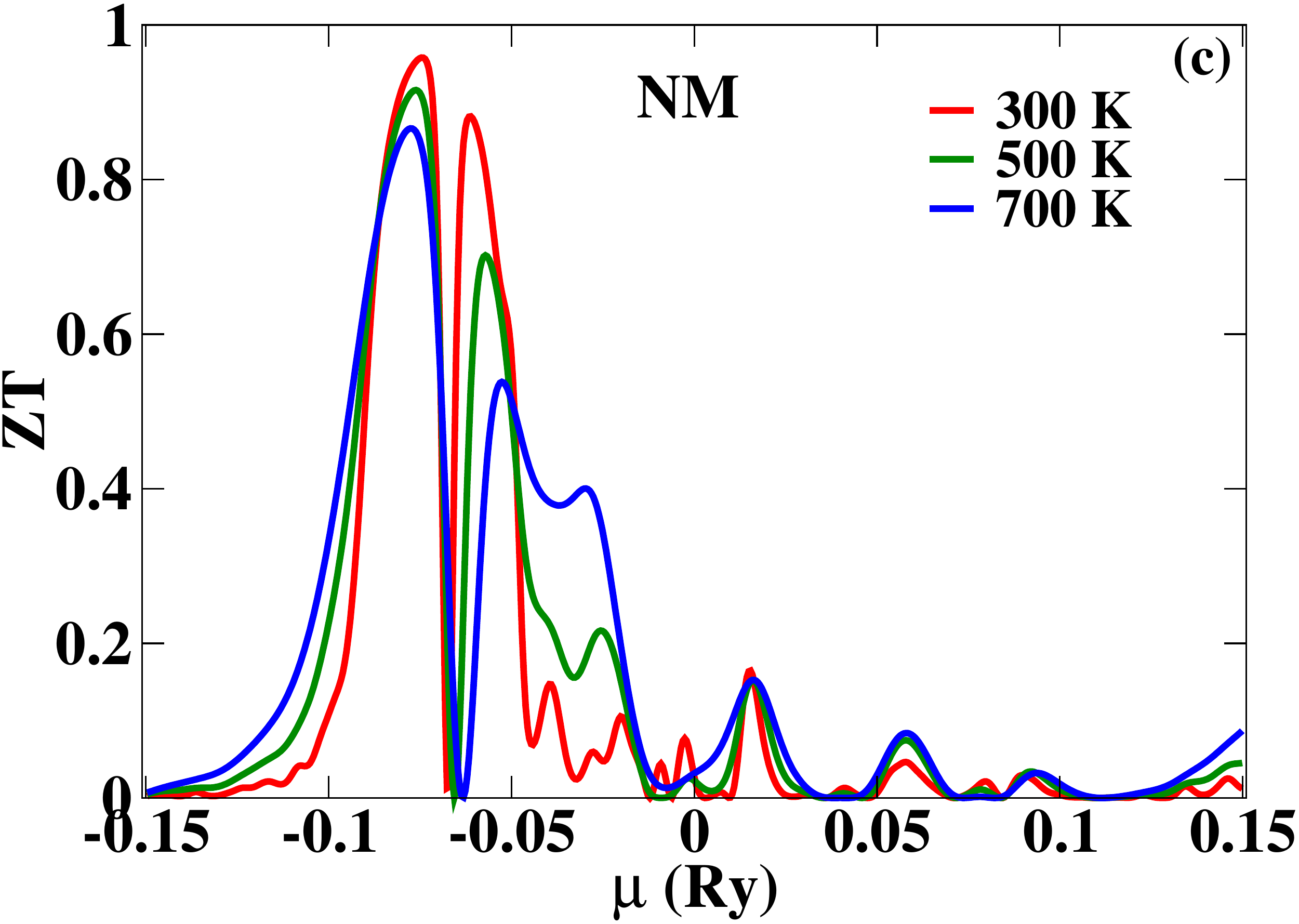}\includegraphics[scale=0.3]{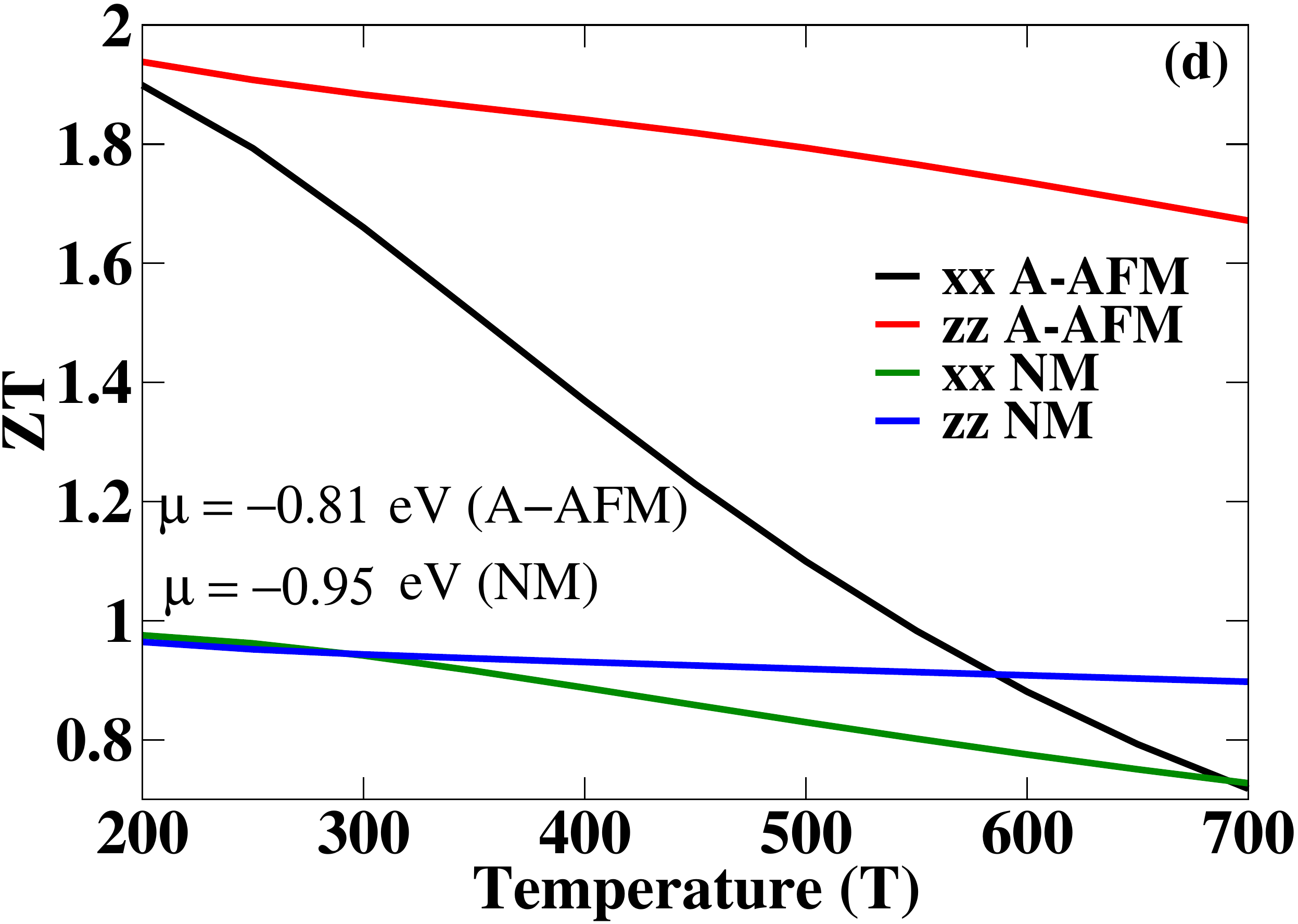}
	\caption{(a) Variation of average power factor ($S^{2}\sigma$) w.r.t. temperature for A-type AFM and NM phase along $xx$ and $zz$ direction. 
	 Variation of $ ZT $ w.r.t. $\mu$ for (b) A-AFM, and (c) NM state (The Fermi level is set at $\mu$= 0 eV), (d) temperature dependence of $ZT$  along $xx$ and $zz$ direction for fixed value of $\mu$ = -0.81 eV and -0.95 eV for A-AFM and NM cases respectively. }
\end{figure*}
Fig. 5 shows the electronic part of thermal conductivity ($\kappa_{e}$/$\tau$) as a function of $\mu$ for A-AFM and NM at different temperatures. It is apparent from figure that as $\mu$  increases, the $\kappa_{e}$/$\tau$ also increases till a peak is achieved at 0.95 eV and 1.5 eV in case of A-AFM and NM respectively at 300 K. The magnitude of $\kappa_{e}$ for A-AFM is larger at the $\mu >0$ than its $\mu <0$ counterpart. This suggests that for the hole doped system the $\kappa_{e}$/$\tau$  is less hence its optimal for thermoelectric application. On the other hand we observed slightly higher magnitude of $\kappa_{e}$/$\tau$ along $\mu <0$ for NM case. When we fix $\mu$ at the value where $S$ was maximum, and observe the variation of $\kappa_{e}$/$\tau$ w.r.t. temperature, we observed an increase in $\kappa_{e}$/$\tau$ with the increase in temperature along $xx$ and $zz$ which is because of increase in charge carriers. The $\kappa_{e}$ along $xx$ and $zz$ cases for A-AFM has value 3.9022$\times$$10^{13} $ and 1.8770$\times$$10^{13} $ at 500 K for a fixed $\mu$. The low $\kappa_{e}$ value found along $zz$ direction actually implies high scattering along c-axis in A-AFM case.\\
As stated above, the best performance of the thermoelectric material is judged by proper tuning of its TE parameters at the operating range which can be actually characterized by measuring its power factor (PF). For a fixed values of $\mu$ ($\mu=-0.81$ for A-AFM and $\mu=-0.95$ for NM) we have studied the variation of PF as a function of temperature. The Fig. 6(a) shows the comparison of PFs of A-AFM and NM cases along $xx$ and $zz$ direction as a function of temperature. There is an increase in PF as temperature increases. More importantly, we observe that the PF for the A-AFM becomes significantly higher than NM case above 300K that indicates qualitatively better TE performance in the former. In A-AFM the PF along $xx$ is greater than $zz$ up to 600 K after which the contribution along $zz$ enhances. We have also investigated the figure of merit (ZT) (considering only the electronic part of thermal conductivity). A remarkably high 
value of ZT (between 3 to 4) for A-AFM is observed at $\mu$ $\sim$ -0.5 eV in the temperature range 300K to 700K which is because of the presence of $4f^{7}$ states of the $Eu$ ion. At 300 K for A-AFM, ZT of 0.77 is obtained for p-type doping (i.e. $\mu$ $\sim$ -0.12 eV) and a ZT of 0.93 for n-type doping (i.e. $\mu$ $\sim$ 0.05 eV). For NM case, ZT decreases with increase in temperature and is higher for p-doped region. We obtained ZT of 0.95 at $\mu$ = -0.95 eV for 300 K in NM. With respect to temperature, ZT decreases because of increase in $\kappa_{e}$ as temperature increases. Using $xx$ or $zz$ components of $S$, $\sigma$ and $\kappa_e$, we obtain at 500K, a ZT of 1.09 and 1.79 respectively for A-AFM whereas for NM the ZT along $xx$ is 0.84 and $zz$ is 0.9 (see Fig. 6(d)). These observations suggest that the A-AFM phase will give better TE performance along $zz$ direction. 

\section{Conclusions}
We have performed a detail study of the thermoelectric properties of a 122-type Zintl phase compound namely,  EuCd$_{2}$As$_{2}$ using first principles density functional theory and Boltzmann transport theory calculations to quantitatively estimate the potential of this compound for thermoelectric applications. Previous experimental and theoretical studies have shown an anisotropy in transport and magnetic properties of this compound. First, we performed a detailed analysis of the structural stability for a magnetic A-AFM (the ground state) and non-magnetic (NM) structures using phonon calculations and establish the dynamical stability of the compound in both magnetic and non-magnetic configurations. We then calculated various thermoelectric parameters such as Seebeck coefficient ($S$), electrical and thermal conductivity, power factor and ZT for A-AFM and NM configurations along $xx$ and $zz$ directions. We observe a clear anisotropy in almost all thermoelectric parameters evaluated with A-AFM having stronger anisotropic properties than that in NM. Because of the presence of flat $f$-bands just below Fermi level, we observe very high Seebeck coefficient of 1423 $\mu$V/K and 825 $\mu$V/K for p-doping in A-AFM and NM case respectively. High value of $S$ and low value of $\sigma$ obtained along $zz$ direction for A-AFM phase makes it a better candidate for thermoelectric applications which is reflected in the TE figure of merit $ ZT $ value of 1.79 at 500 K. Even the NM phase has ZT values close to 1. From the variation of $ ZT $ as a function of $\mu$ we observed that with p-type doping the $ ZT $ is enhanced appreciably. Hence, we propose  EuCd$_{2}$As$_{2}$ to have a very good potential for application as a TE material.
\section{Acknowledgement} TM and JK acknowledge useful discussions with S. Auluck. JK acknowledges research fellowship from MHRD, India.

\end{document}